\documentclass[pre, twocolumn, showpacs, superscriptaddress, floatfix]{revtex4}

\usepackage{amssymb, amsmath, amsfonts}
\usepackage{bm}
\usepackage[dvips]{graphicx}
\usepackage{subfigure}

\begin{document}

\title{Feedback topology and XOR-dynamics in Boolean networks with 
varying input structure} 

\date{May 28, 2009}

\author{L. Ciandrini}
\email{l.ciandrini@abdn.ac.uk}
\affiliation{Universit\`a di Pavia, Dip. di Fisica Nucleare e Teorica,
  Via Bassi 6,  27100 Pavia, Italy}
\altaffiliation[Present address: ]{Institute for Complex Systems and Mathematical Biology, KingÕs College,
University of Aberdeen, United Kingdom}
\author{C. Maffi}
\email{carlo.maffi@epfl.ch}
\affiliation{Laboratoire de Biophysique Statistique, EPFL SB ITP,
  CH-1015, Lausanne, Switzerland}
\author{A. Motta}
\affiliation{Universit\`a degli Studi di Milano, Dip. Fisica, Via
  Celoria 16, 20133 Milano, Italy}
\author{B. Bassetti}
\affiliation{Universit\`a degli Studi di Milano, Dip. Fisica, Via
  Celoria 16, 20133 Milano, Italy}
\affiliation{I.N.F.N. Milano, Italy}
\author{M. \surname{Cosentino Lagomarsino}}
\email{Marco.Cosentino-Lagomarsino@unimi.it}
\affiliation{Universit\`a degli Studi di Milano, Dip. Fisica, Via
  Celoria 16, 20133 Milano, Italy}
\affiliation{I.N.F.N. Milano, Italy}


\begin{abstract}
  We analyse a model of fixed in-degree Random Boolean Networks in
  which the fraction of input-receiving nodes is controlled by the
  parameter $\gamma$. We investigate analytically and numerically the
  dynamics of graphs under a parallel \textsc{xor} updating
  scheme. This scheme is interesting because it is accessible
  analytically and its phenomenology is at the same time under
  control, and as rich as the one of general Boolean networks.
  We give analytical formulas for the dynamics on general graphs,
  showing that with a \textsc{xor}-type evolution rule, dynamic
  features are direct consequences of the topological feedback
  structure, in analogy with the role of relevant components in
  Kauffman networks. Considering graphs with fixed in-degree, we
  characterize analytically and numerically the feedback regions using
  graph decimation algorithms (Leaf Removal). With varying $ \gamma$,
  this graph ensemble shows a phase transition that separates a
  tree-like graph region from one in which feedback components emerge.
  Networks near the transition point have feedback components made of
  disjoint loops, in which each node has exactly one incoming and one
  outgoing link. Using this fact we provide analytical estimates of
  the maximum period starting from topological considerations.
\end{abstract}

\maketitle

\section{Introduction}
Biological networks are graphs representing the basic interactions
between molecules in a living cell \cite{barabasi:2004aa, alon:2003aa,
  hartwell:1999aa, bray:2003aa}. They are generally composed of a
fairly large number of elements and for this reason it is unfeasible
to treat all the biochemical processes in detail. For example, a
transcription network \cite{oltvai:2002aa, lee:2002aa, babu:2004aa} of
a simple cell counts some thousands of nodes, which represent the
genes of a cell. Given the interaction structure, it is important to
establish which genes are active in a given time or environment, or
under a given stimulus.  In order to fulfill this task, it is
necessary to develop coarse-grained models for gene expression, such
as discrete systems, in which variables on the nodes represent the
(discretized) expression of single genes, and directed connections
stand for their interactions \cite{thomas:1973aa, kauffman:1993aa}. These 
abstract models are useful to study on general grounds the emergent cooperative behaviour of gene
expression, as biological function is increasingly being recognized as 
emerging from global phenomena rather than from the expression of
single genes \cite{barabasi:2004aa, guelzim-n:2002aa, isalan-m:2008aa}.

The simplest model of this kind are Random Boolean Networks (RBNs)
introduced by S. Kauffman in 1969 \cite{kauffman:1969aa}. In this
model, $N$ elements take binary values and interact with some random
coupling functions. In the standard Kauffman model, the configuration
of an element is set by Boolean functions whose values depend on a
fixed number $k$ of inputs. The system is specified by its topology (a
graph with $k$ inputs regulating each node), a synchronous updating
scheme and the choice for the ensemble of Boolean functions (for an
introductory review see Refs. \cite{gershenson:2004aa, aldana:2003ab,
  drossel:aa}).
Technically speaking, the behaviour of the system is fully
characterized by its cycles (or fixed points) and their basins of
attraction. If a cycle contains exponentially many (in $N$)
configurations, the behaviour is called chaotic. Otherwise it is
called ordered. According to the original Kauffman interpretation, if
the system is in a chaotic state, it cannot exhibit specific behaviour
in response to external stimuli. More specifically, a realization of
the model is interpreted as a genome and an attractor as a possible
cell type (this particular interpretation is today considered out of
date \cite{Ribeiro:2007rm}). If the corresponding cycle period is too
long, this hypothetical cell type would never realize it. For this
reason, networks of biological interest should lie between the ordered
and the chaotic phase (the so-called critical networks
\cite{bastolla:1997aa}), where attractor cycles are neither too short
nor exponential.\\
\indent There are two problems concerning Kauffman's model in relation with
genetic networks. First, the ensemble contains only graphs entirely
made of feedback, while typically this is not the case of biological
(e.g. transcription) networks, which usually have some ``sensor''
nodes that respond to external conditions \cite{seshasayee:2006aa,
  luscombe:2004aa, balaji:2007aa}. Thus, it is useful to modify the
model and consider networks with well-defined input structure. The
second problem is connected to the choice of the ensemble of Boolean
functions. While on biological grounds it is difficult to characterize
this class, from a mathematical viewpoint the choice of functions
strongly conditions the behaviour of the model. Indeed, since most
functions are constant, or ``canalizing'' \cite{aldana:2003ab,
  drossel:aa} with respect to some variables, this creates an
``effective topology'', which does not correspond to the underlying
interaction graph.  For this reason, the study of attractors in
Kauffman networks is complex (for recent results see, for example,
\cite{paul:2006aa, samuelsson: 2003aa, drossel:2005ab}). Also the
study of critical networks with canalizing Boolean functions leads to
the conclusion that this choice does not decrease attractor sizes and
render the model more well-behaved as previously expected
\cite{paul:2006aa}.

In this work, we address these questions in a controlled way, with the
following choices.  First, we work with a graph ensemble with a
parameter that regulates the fraction of input nodes, as done in
prvious studies that focused on fixed points \cite{
  cosentino-lagomarsino:2005aa, correale:2006aa, correale:2006ab,
  cosentino-lagomarsino:2006aa}.  Secondly, we choose to use the
ensemble of \textsc{xor} or totally non- canalizing functions.
In other words, we consider the situation in which all the existing
links in the network have a role in the dynamics of real systems.  The
aim is to study the repercussions of the chosen topology on the
dynamics and how this restricted selection of Boolean functions, in
which the output changes if any one of the inputs changes its value,
strictly relates these two aspects of the model.  From the technical
viewpoint, this dynamics has the advantage to be approachable with
linear algebra using the finite Galois field $GF(2)$
\cite{kolchin:1998aa, caracciolo: 2002aa}, the set $\{0,1\}$ where
addition between elements is equivalent to the logic \textsc{xor} and
multiplication to the logic \textsc{and}.

Our main results are the following. With varying $\gamma$, we observe
a phase transition, similar to the one observed in Kauffman networks,
from a region characterized by an ordered dynamic behaviour to a
region in which the dynamics is chaotic. From a topological point of
view the same transition divides a region characterized by tree-like
graphs from one in which extensive feedback is present. The structure
of networks around the critical point appears simplified, in the sense
that the feedback regions of this kind of networks are organized in
simple disjoint loops. We use the topological structure of critical
point networks to estimate the maximum period times. The paper is
organized as follows. After giving the basic definitions in
Section~\ref{sec:model}, we present the main results in Section~
\ref{sec:results}. In the last sections we discuss the results 
and the parallels with the scaling approach to general Kauffman
networks~\cite{mihaljev:2006aa}.
\section{Definition of the model\label{sec:model}}
\begin{figure}[htbp!]
  \begin{center}
     \includegraphics[scale=0.4]{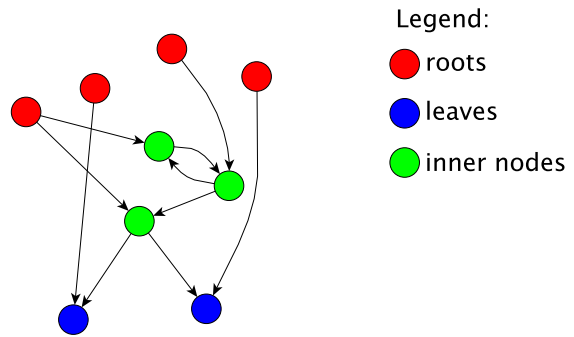}
     \caption{A network with $N=9$, in-degree $k=2$ and $\gamma \simeq
       0.55$. Red circles (colors online) indicate roots, blue circles leaves 
and
       green circles the inner nodes.\label{fig:fig1}} 
  \end{center}
\end{figure}

We consider networks of $N$ nodes, of which only $M$ receive input,
and define $\gamma = M/N$. Each of these regulated nodes receives
inputs from exactly $k$ randomly and independently chosen other
nodes. Consequently the in-degree is $k$ or $0$, while the out-degree
distribution is binomial, and in the ``thermodynamic'' limit $N
\rightarrow \infty$ and $\gamma$ constant is well approximated by a
Poisson distribution of mean $k\gamma$,
\begin{equation*}
	p_{out}(\gamma k, n) = e^{-\gamma k}\frac{(\gamma k)^n}{n!} .
\end{equation*}
As a consequence, the graphs contain nodes with no output
(\textit{leaves}) and $N-M$ \textit{roots}, i.e. variables without inputs
(Fig.~\ref{fig:fig1}). This ensemble is conventionally used in
diluted spin-glass models and constraint-satisfaction networks 
\cite{mezard:aa, mezard:2003aa}. The standard Kauffman graph 
ensemble
can be obtained considering the special case $\gamma = 1$ in which all
variables are regulated.
The graph ensemble is specified by a $M \times N$ connectivity matrix
$A$ where $a_{ij}=1$ if the edge $j\rightarrow i$ exists, and zero
otherwise. It is useful to arrange the columns of $A$ in order to
reserve the first $M$ indices to the nodes that receive inputs. This
matrix can further be divided into two submatrices $A = ( S \; | \;
R)$. $S$ is the $M \times M$ matrix that describes the interaction
between regulated elements, while the columns of $R$ contain the
information on the outputs of the free variables.

We consider a dynamics on the graphs of this ensemble, specified by
assigning Boolean variables to each of the $N$ nodes and interactions
through \textsc{xor} coupling functions. A global state $\Vec{\sigma}$
is defined as the set of configurations assumed by all the nodes of
the network. The initial conditions determine the state of the root nodes 
since they do not have input 
and their values remain fixed 
during the evolution. Root nodes are considered as external inputs and for 
this reason we can think 
that the initial conditions 
represent ``the external world''. The configuration space, formed by all 
global states, contains $2^N$ 
states. Since the system is 
finite, starting from some
initial global state, the deterministic dynamics leads to periodically
repeated states, possibly after a transient time. In other words, the
system performs a trajectory in the state space and eventually arrives
to an \textit{attractor} of length $T$, where the states are periodic in time. 
The attractor has length $T$ if
$\Vec{\sigma}(t+T) = \Vec{\sigma}(t)$. We call this a $T$-cycle or a
\textit{fixed point}, in case of unitary length. The \textit{basin} of
attraction of an attractor is the set of global states that reach the
attractor, including the attractor states. A \textit{transient state} is
a state that belongs to a basin but is not part of an attractor. The
$N$-dimensional Boolean vector $\Vec{\sigma}(t)$ can be written as
$\Vec{\sigma}(t) = (\Vec{x}(t) \, , \, \Vec{y}(t))$. The
$M$-dimensional vector $\Vec{x}(t)$ denotes the state of the regulated
variables and the ($N-M$)-dimensional vector $\Vec{y}(t)$ the
(constant) state of the root variables ($\Vec{y}(t) = \Vec{y}(0) =
\Vec{y}$).\\
\indent The synchronous update at time $t$ is determined by random
\textsc{xor} functions, or, equivalently, by the linear operation in
the Galois field $GF(2)$ \cite{kolchin:1998aa,
  cosentino-lagomarsino:2006aa}
\begin{equation}
   \label{xor}
   \Vec{x}(t+1) = A \Vec{\sigma}(t) + \Vec{b} \; = S \Vec{x}(t) + R
   \Vec{y} + \Vec{b}, 
\end{equation}
where $\Vec{b}$ is a random $M$-dimensional Boolean vector containing
the information relative to the nature of regulating functions. The
components $b_i$ of $\Vec{b}$ are $0$ or $1$ with probability
$\frac{1}{2}$. The evolved state after $t$ steps is 
\begin{equation}
   \label{x(t)}
   \Vec{x}(t) = S^t \Vec{x}(0) + \sum_{i=0}^{t-1}
   S^i(\,R\Vec{y}+\Vec{b}\,) = S^t \Vec{x}(0) +\Sigma(t)(\,R
   \Vec{y}+\Vec{b} \,) , 
\end{equation}
where $\Sigma(t) \doteq \sum_{i=0}^{t-1} S^i$ (the symbol $\doteq$
stands for a definition). One can observe from Eq.~(\ref{x(t)}) that
the dynamics is controlled by the topology through the interaction
matrix. The peculiarity of the \textsc{xor} dynamics is that every
change of one (or an odd number of) input in a function determines a
change in the output.  We shall demonstrate that feedback components
of the graph are the only relevant region needed for characterizing
the dynamic behaviour of the whole network.\\
\indent The model introduced here bears some similarities and some differences
with the Kauffman model. One difference is that the graph ensemble is
not the same, as only a fraction of nodes receives input.  However, as
we will see, root nodes and feedback regions play the same role of
nodes with a constant update function and so-called ``relevant'' nodes
\cite{flyvbjerg:1988aa, bilke:2001aa} respectively, leading to an
interesting parallel with more general Boolean
networks~\cite{mihaljev:2006aa}.  In the standard Kauffman model, the
existence of frozen nodes (i.e. nodes under constant update functions,
where the variables assume the same value on every attractor) denotes
that some links are irrelevant to the dynamics and effectively
modifies the topology in a way that is, in general, difficult to
control.  By contrast, the simplified model presented here enables to
separate the discussion and the study of topology and dynamics, the
features of the dynamics being a repercussion of the topology of the
network. Note however that in our case the value of root nodes can
change with the initial conditions in contrast with what happens to
frozen nodes in Kauffman networks.

\section{Results\label{sec:results}}
\subsection{General features of XOR-dynamics \label{sec:dyn}}
The discrete \textsc{xor} dynamics defined above allows to derive some
simple general properties that will be used in the following.

Firstly, linearity implies that the cycles have a least common
multiple (\textsc{lcm}) structure, where longer nontrivial cycles can
be constructed by combining smaller ones. As a consequence, if a
network shows a set $\{T_i\}$ of cycle lengths, then a cycle of length
$T' = \mathrm{lcm} \{T_i\}$ also exists.

From Eq.~(\ref{x(t)}) a global state $\Vec{\sigma}(t)$ belongs to
a $T$-cycle if $ \Vec{\sigma}(T+t)=\Vec{\sigma}(t) $, which gives the
condition
\begin{equation*}
  \left[
    (1+S)\Vec{x}(0) + R\Vec{y} + \Vec{b}\right] \in \ker\Sigma(T)\; , 
\end{equation*}
i.e. the vector on the left hand side is sent to $\Vec{0}$ by the
function $\Sigma$ for some $T$. In the eventuality of $T=1$ the
same condition becomes
\begin{equation*}
  (1+S)\Vec{x}(0) =  R\Vec{y} + \Vec{b}  \;. 
\end{equation*}

The fact that the dynamics can only have transients or cycles
translates into the formula
\begin{equation}
  S^{l+t_{m}} = S^{l} \;,
  \label{cycl}
\end{equation}
where $l$ is the smallest integer such as $\ker S^l = \ker S^{l+1}$
(the length of the longest transient) and $t_{m}$ is the maximum cycle
length (without repetitions) for the map $\Vec{x} \rightarrow
S\Vec{x}$. 
Indeed, after a transient period of at most $l$ steps, the dynamics
becomes cyclic, i.e. identical configurations occur each $t_{m}$ steps
(and $S^{t_m}$ is the identity matrix).
This quantity can be related to the maximum cycle for the
complete dynamics using the fact that $\Sigma(2t_m + l) = \Sigma(l)$,
so that for any given initial condition,
\begin{equation*}
   \Vec{x}(2t_m + l) = S^l \Vec{x} + \Sigma(l)(\,R \Vec{y} + \Vec{b})
   = \Vec{x}(l) \; ,
\end{equation*}
which implies that the maximum cycle for the full dynamics is at most
$T_{max}= 2 t_m$.

We now give a heuristic argument connecting the algebraic properties
of $S$ with statistical properties of the dynamics.  If $m(T)$ is the
multiplicity of $T$-cycles, and $\mathcal{B}(T)$ is the number of
configurations in the basin of attraction of each $T$-cycle, one must
have
\begin{equation*}
  2^N = \sum_T \mathcal{B}(T) m(T) \;.
\end{equation*}
The elements belonging to a basin of attraction are built adding
$\Vec{z} \in \ker S^l$ to a fixed system state $\Vec{x}$ of a
$T$-cycle (Note that also elements $\Vec{z} = \Vec{0}$ are taken in
consideration and thus $\Vec{x}$ itself is considered part of the
basin). Thus, each attractor of length $T$ has a basin of attraction
whose dimension is $\mathcal{B}(T) = T \mathcal{B}(1) = T\,2^d$, where
$d = \dim(\ker S^l)$.
To compute the probability of a cycle one has to provide an estimate
for $m(T)$.  From previous work \cite{cosentino-lagomarsino:2005aa},
we know that generally the average number of fixed points is $m(1) =
2^{N-M}$, this can be also the typical value, depending on the graph
ensemble and the value of $\gamma$.  Since adding a fixed point to the
elements of a $T$-cycle leads to a cycle of the same length, $m(T)$
has a prefactor $m(1)$. Supposing that the contribution of the longer
cycles is of order one, one can write, $m(T)\sim m(1)$, or in other
words the multiplicity of cycles can be estimated using the number of
fixed points.
In this hypothesis, the probability $p(T)$ that a state is in the
basin of attraction of a $T$-cycle is
\begin{equation}
	\label{pT}
	p(T) = \frac{1}{2^N}\mathcal{B}(T) m(T) \simeq 2^{d-M}T \;.
\end{equation}
Note that this estimate holds only for the values of $T$ that are
possible. If $T_{max}$ grows faster than $M$, $\sum_T T \simeq
T_{max}$ implying that $P(T)$ will be concentrated on $T_{max}$.
Equation (\ref{pT}) has a practical implication for simulations: given
a realization of a network, it enables to simulate only few initial
conditions to find the value of $T_{max}$ that is reached with highest
probability. These facts will help understanding some dynamic features of
the networks around the critical line.

\subsection{Simulations}
We will now turn our attention to direct simulations of the model
presented in Section~\ref{sec:model}.  In these simulations one has to
sample a number of ensemble graphs, and for each graph, a number of
random initial conditions.  The choice of the average performed and
the quantity of interest leads to the definition of different
observables.

\begin{figure}[htbp!]
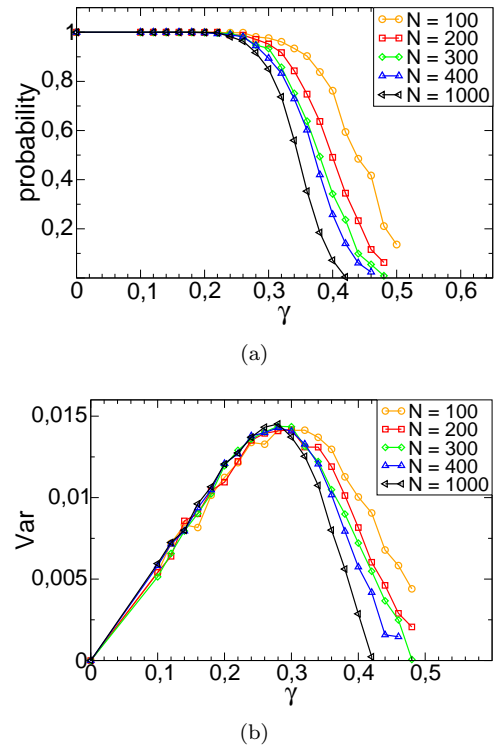

\centering
\subfigure[\label{fpoints}]%
{\includegraphics[width=0.75\linewidth]{fig2_a.eps}}\\ 
\subfigure[\label{varfpoints}]%
{\includegraphics[width=0.75\linewidth]{fig2_b.eps}}
\caption{\textbf{Simulations of dynamics: Fixed points.} (a)
  Probability of finding at least one fixed point for different
  values of $N$ ($k=3$). The probability drops down after a certain
  value of $\gamma$. (b) Variance of the probability of observing
  one fixed point as a function of $\gamma$. The plot is obtained by
  averaging the fraction of $10^3$ random initial configurations
  reaching a fixed point (on a fixed graph) over $10^3$ graph
  realizations at a given $\gamma$. \label{old}}
\end{figure}

The simplest observables relate to the fixed points of the dynamics.
Figure~\ref{old}.a shows the probability of finding at least one fixed
point as a function of the parameter $\gamma$. Fixed points are
typically exhibited for low values of $\gamma$ while they become rare
for higher values, with a sharp threshold in between. One can notice
that increasing system size makes more evident the two distinct
regimes.
Another way of presenting this result is given in Fig.~\ref{old}.b,
which shows the variance of $p(1)$ depending on $\gamma$. This plot
has a peak whose maximum value shifts with increasing number of
nodes. Accurate location of the transition point in Figs.~ \ref{old}.a
and \ref{old}.b is rendered difficult by the fact that for larger
system sizes, where the results should be more reliable, sampling
efficiently the initial conditions becomes difficult.
\\
\begin{figure}[!ht]
 \centering
 \includegraphics[width=0.8\linewidth]{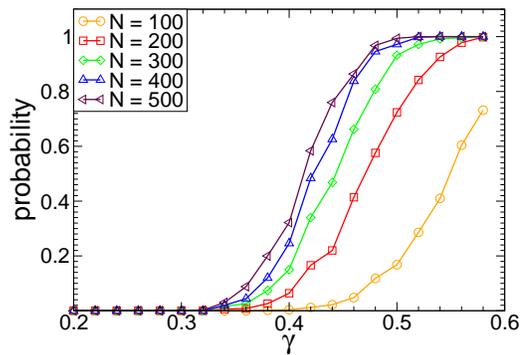}
 \caption{\textbf{Simulations of dynamics: Long cycles.} The plot
   presents the fraction of periods larger than the cutoff
   ($10^6$). At $N$ and $\gamma$ given, we sampled $10^{3}$ random
   networks and for each network we evaluated $10$ initial
   conditions.  \label{over_cf}}
\end{figure}

These results show signatures of a dynamical transition point $
\gamma_c^d$ which marks the border between a region with typically
fixed points and a region where cycles dominate on fixed points (for
recent studies on the critical transition between chaotic and ordered
phase in Kauffman networks see, e.g.,
\cite{Andrecut:2008ty}). According to Eq.~ \ref{pT}, for $T=1$, the
shape of the curves indicates that $d$ becomes significantly different
from $M$ after a certain value of $ \gamma$. In the large system
limit, if $d \ll M$ one expects that $p(1)$ is negligible and the
probability of finding at least one fixed point drops to zero. At the
transition point the probability of a fixed point and its variance
should drop to zero in the thermodynamic limit. The effects of finite
size are evident as the measured transition point strongly depends on
the system dimension. We will investigate these effects later
(Section~\ref{sec:LeafRemoval}). Note that the fraction of fixed
points just below the threshold decreases unexpectedly with $N$.
However, this has no physical meaning and is connected to the fact
that the number of initial conditions is kept constant and
consequently larger systems are increasingly under-sampled.

The same transition is visible by quantifying the length of
cycles. However this task is computationally hard already for $N
\simeq 100$, which strongly limits this kind of simulations. Indeed,
to implement these simulations, it is necessary to impose a cutoff on
the length of the maximum period observed. Figure~\ref{over_cf} shows
the fraction of networks with a period larger than the cutoff imposed,
given $N$ and $\gamma$. This quantity grows rapidly beyond a
characteristic parameter value that changes even more with the system
dimension, and the transition point is difficult to locate
precisely. Hence, it is necessary to find effective methods to
investigate the dynamics and this problem will be discussed in
Section~ \ref{sec:dynamic_core} after having studied the feedback
topology of graphs in correspondence with the transition.

\subsection{Feedback topology and Leaf Removal algorithms 
\label{sec:LeafRemoval}}
In this section, we analyze the topology of the graphs.  The question
that we want to address is whether the dynamic transition is connected
to a change in the feedback topology of the underlying graphs.

To study the feedback regions we use three modified versions for
directed graphs of the Leaf Removal (LR) decimation
algorithm \cite{mezard:2003aa, bauer:2001aa}. A graph decimation
technique consists in erasing iteratively nodes and links with some
specified prescriptions. The variants we implemented remove tree-like
parts of the graph, leaving the components with feedback. Network
regions not removed by the algorithms are called \textit{feedback core
  of the graph} or simply \textit{core}.  It is possible to study
analytically the behaviour of the algorithms in the mean-field
approximation \cite{weigt:2002aa}, considering for example the
fraction of nodes in the core.  The LR algorithms define a dynamics
for the probability $p_n(t)$ that a regulated node has $n$ inputs at
iteration $t$, (i.e. of having $n$ ones on a row of the matrix $S$),
and for the probability $f_{n}(t)$ that a node has $n$ outputs at
iteration $t$ ($n$ ones on a column of $S$).  We study LR algorithms
analytically with mean-field equations and compare with numerical
simulations.
As we will discuss, this approach is related to the scaling approach~
\cite{mihaljev:2006aa} to general Kauffman networks, where one can
write and analyze scaling equations for the numbers and fluctuations
of nodes receiving input from a given number of frozen nodes.

\begin{figure}[htbp!] 
	\centering
    	\includegraphics[width=0.718\linewidth]{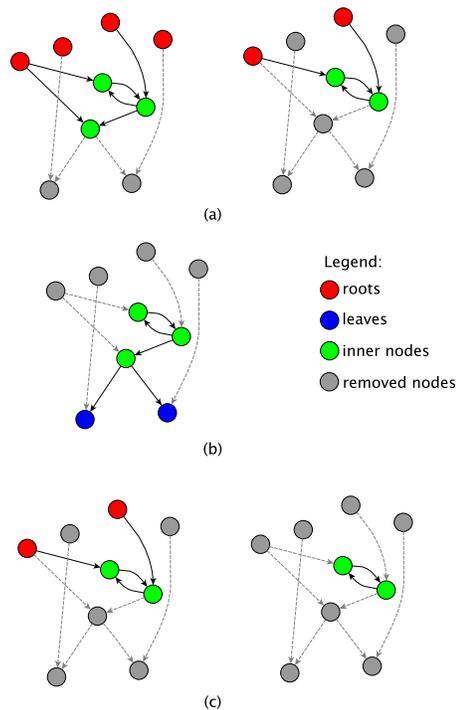}
    \caption{\textbf{Leaf Removal algorithms}. (a) Iterations of LRu
      for the graph presented in Fig.~\ref{fig:fig1} . The LRu removes
      iteratively the non regulating variables and their links (light grey
      nodes and dotted arrows). In this case the algorithm stops after
      two steps. (b) Iterations of LRd. The LRd removes iteratively
      the regulating variables that are not regulated and the
      associated links (light grey nodes and dotted arrows). Only one step
      is possible in this case. (c) Iterations of LRb. First the LRu
      is applied and then all non-regulating nodes are taken off with
      the LRd. \label{Leaf Removal algo}}
\end{figure}

\begin{figure}[htbp]
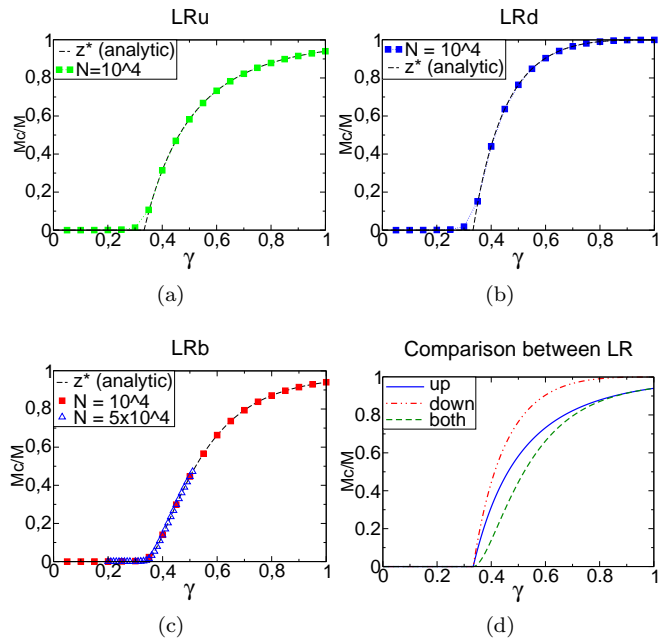
 
	\centering%
	\subfigure[\label{LR:a}]%
        {\includegraphics[width=0.494\linewidth]{fig5_a.eps}}
	\subfigure[\label{LR:b}]%
        {\includegraphics[width=0.494\linewidth]{fig5_b.eps}}\\
        \subfigure[\label{image_down:a}]%
        {\includegraphics[width=0.495\linewidth]{fig5_c.eps}}
	\subfigure[\label{image_both:b}]%
        {\includegraphics[width=0.493\linewidth]{fig5_d.eps}}
        \caption{\textbf{Phase transition for LR algorithms}
          ($k=3$). The dashed lines represent the fraction of nodes
          removed by LR up (a), LR down (b) and LR both (c) calculated
          for different values of $\gamma$. Squares are numerical
          results for graphs with $N=10^4$ nodes and blue triangles in
          (c) are the simulations for $N=5 \cdot 10^4$. Each point is
          obtained averaging $10^4$ simulations. Error bars are
          smaller than the dimension of the points. The discontinuity
          of the derivative points to the emergence of an extensive
          feedback core at $\gamma = \frac{1}{3}$. (d) Comparison of
          analytic curves for the three variants of the
          algorithm.\label{LR}}
\end{figure}

The first LR variant we analyze is the so called \textit{Leaf Removal
  Up} (LRu) \cite{cosentino-lagomarsino:2006aa}. In each step of this
algorithm the variables without outgoing links (the leaves) are
removed together with its incoming links. The procedure is iterated
until each variable has at least one outgoing edge (Fig.~\ref{Leaf Removal
  algo}.a).
As a consequence of the removal of links, the distribution $f_n$ of
the outgoing links changes after each iteration $t$, giving rise to
fluxes that can be expressed by first-order mean-field equations for
$f_n(t)$ (see Ref. \cite{cosentino-lagomarsino:2006aa} and Appendix~
\ref{sec:appA}). The solution of the mean-field flux equations is:
\begin{equation*}
	 \begin{cases}
	f_{0}(\bar t\,) = -\bar{t}+e^{-\lambda(\bar{t} \,)} \\
	f_{n}(\bar t\,)  =   e^{- \lambda(\bar t\,)}\frac{\lambda (\bar t\,)^{n}}{n!} \;,
\qquad n>0 \;,
	\label{sol_up}
	\end{cases}
\end{equation*}
where $\bar t = t/M$ and $\lambda(\bar t\,) = \gamma k (1- \bar t\,)$.
The algorithm stops at the reduced time $\bar{t}^*_{up}$
and the number $M_c^{up}$ of remaining nodes is
\begin{equation*}
	M_c^{up} = Mz^*_{up} \;,
\end{equation*}
where $z^*_{up} = 1- \bar t^*_{up} $ is a function of $\gamma$ and
represents the fraction of nodes belonging to the LRu core on the
total of regulated nodes.  Figure~\ref{LR}.a compares the analytic curve for
$z^*_{up}$ as a function of $\gamma$ with numerical results. Varying
$\gamma$, there is a singularity which is a signature of a phase
transition between a region characterized by graphs entirely removed
by the algorithm ($z^*_{up}=0$) and a region in which the amount of
remaining nodes after the process augments when $\gamma$ is
increased. These regimes are separated by the ``critical point''
$\gamma_c^{up} = k^{-1}$ found with mean field equations.

The second variant we analyze is the \textit{Leaf Removal down}
algorithm (LRd) \cite{maffi:2005aa}, which, in a similar way as
above, removes iteratively the nodes with no incoming links, together
with their outgoing edges, until there are no more roots (see
Fig.~\ref{Leaf Removal algo}.b). The mean-field flux equations (shown in 
Appendix~\ref{sec:appA}) for the distribution of incoming links
$p_n(t)$ have solution
\begin{equation*}
 	\begin{cases}
	p_0(z) = (z-1)+(1-\gamma z)^k\\
	p_{n}(z)=\left(\begin{array}{c}k \\ n\end{array}\right) (\gamma z)^n (1-
\gamma z)^{k-n} \;, \qquad 
n>0 \;,
	\end{cases}
	\label{sol_down}
\end{equation*}
where $z=1-\bar t$.  When the algorithm halts the amount of nodes in
the LRd core is given by
\begin{equation*}
	M_c^{down} = Mz^*_{down} \;.
\end{equation*}
The critical value $\gamma_c^{down} = k^{-1}$ divides the networks
with a non-vanishing core in the large $N$ limit from the ones that
are removed by LRd. The plot of $z^*_{down}(\gamma)$ is presented in
Fig.~\ref{LR}.b and compared with simulations.

The last version of LR we consider is the combination of LRu and LRd
which we call \textit{Leaf Removal both} (LRb).  From the point of view
of the dynamics this algorithm first removes all the nodes which do
not regulate the \textit{core} and successively it removes the set of
nodes that are fixed, at most after a transient time, the initial
conditions given.  From the topological viewpoint it leaves all and
only the nodes involved in feedback cycles (Fig.~\ref{Leaf Removal
  algo}.c).  It can be checked that a single iteration of LRu and LRd
is sufficient for this.  The LRu algorithm stops at a matrix with
$M_c^{up}$ rows having $k$ entries and whose remaining rows are
empty. When LRd is applied to this matrix, elimination of all empty
rows and columns leaves with $M_c^{both} = M_c^{up}z^*_{down}=
Mz^*_{down}z^*_{up} $ nodes, and the following per-row and per-column
probabilities of $n$ entries
\begin{equation}
	\begin{cases}
	p_0 = f_0 = 0 \\
   		 p_{n}= (z^*_{down})^{-1} \left(\begin{array}{c}k \\ n\end{array}
\right) (\gamma z^*_{down})^n 
(1-\gamma z^*_{down})^{k-
n}\\
		 f_n =  (z^*_{up})^{-1} e^{- \lambda(\bar t^*_{up})}\frac{\lambda 
(\bar t^*_{up})^{n}}{n!} \;.
	\label{prob_both}
	\end{cases}
\end{equation}
After the application of the algorithm, the distribution of both the
outgoing and the incoming links has changed with respect to the
initial network.  In particular, no nodes without incoming/outgoing
links may remain.
In this case $z^*_{both} \doteq z^*_{down}z^*_{up}$ and $M_c \doteq
 M_c^{both} = Mz^*_{both}$ The core becomes extensive (order $N$)
 above the critical value $\gamma_c^{both} = k^{-1}$ (Fig.~\ref{LR}.c).

The analytical approach described so far evaluates the mean value of
the number of nodes in the core, in the thermodynamic limit $N
\rightarrow \infty$.  In order to access the fluctuations around this
value we use direct simulation. The shape of the distributions changes
significantly with varying $\gamma$ and number of elements.
Figure~\ref{FeedbackCore}.a shows the distributions of $M_c$ with 
varying
$\gamma$. Below the critical value, $M_c$ is condensed around zero,
i.e. the graphs are typically close to being tree-like. Near the
transition, the distributions have broad tails, and for $\gamma \gg
\gamma_c$ they become symmetric around the value found with mean-
field
LR equations. This phenomenology is typical of phase transitions. In
brief, the results show the presence of a phase transition, with
varying the order parameter $\gamma$, between a region of typically
tree-like graphs and a region of extensive feedback loops.  These
conclusions hold independently from the in-degree value $k$.
Figure~\ref{FeedbackCore}.c shows the consequence of finite sizes on the 
LRb
algorithm on the transition: the peak of the variance plot indicating
the transition point is shifted to greater values and, increasing the
system dimensions, it slowly approaches the value $k^{-1}$. Even for
systems with more than $10^4$ nodes, the critical point does not reach
the analytically determined value. We call $\gamma_e$ the maximum of
the variance of $M_c$ i.e. the effective critical value of $\gamma$ at
finite system size ($\gamma_e \longrightarrow \gamma_c, N \rightarrow
\infty)$.
In the region $\gamma_c \leqslant \gamma \leqslant \gamma_e$, a
residual LRb core is observed and, as we will see, this affects the
dynamics.  Thus, for small networks, $\gamma_e$ does not distinctly
separate the region characterized by tree-like graphs and the one
defined by feedback regions which are present also before this value.
This can be also observed in Fig.~\ref{FeedbackCore}.b which presents 
the
scaling of $M_c / N$. At a value of $\gamma = 0.35$, $M_c$ may seem to
be a sub-extensive quantity. However, increasing $N$ it appears clear
that the core is extensive, as the LR equations predict. Taking
simulations of very large systems (up to $5\cdot10^5$ nodes), values
of the fraction of remaining nodes in the core are comparable with
which ones obtained from the analytic calculation.

 \subsection{Feedback structure at the transition}
 Let us now take a closer look at what happens around the critical
 point of LR algorithms.  Figure~\ref{LR}.d shows that the three variants of
 LR have a different trend just after the transition value.  Defining
 $\epsilon = \gamma - \gamma_c$, for small $\epsilon>0$ the rescaled
 times of arrest $z$ (i.e. the solutions of the equations
 $f_0(z^*_{up})=0$ and $p_0(z^*_{down})=0$, see Appendix~
\ref{sec:appA}) become
\begin{eqnarray*}
	z^*_{up} & \simeq & 2k \epsilon(1-2k\epsilon) \simeq 2k\epsilon \\
	z^*_{down} & \simeq & \frac{ 2 \epsilon k^2}{k-1} (1-2 \epsilon k) 
\simeq \frac{ 2 \epsilon k^2}{k-1}
\end{eqnarray*}
and for LRb: 
\begin{equation*}
	z^*_{both} = z^*_{up} z^*_{down} \simeq  \frac{4k^3}{k-1} \epsilon^2 
=  \frac{4k^3}{k-1} (\gamma - 
\gamma_c)^2 \;.
\end{equation*}
\noindent These functions are continuous at the transition but not all
their derivatives are.  The discontinuity in $z^*_{both}$ is of
greater order than the others, signature of transitions of different
orders.
\\
\begin{figure*}[htbp!] 
  \centering%
  \subfigure[\label{Mcdistr}]%
  {\includegraphics[width=0.53\linewidth]{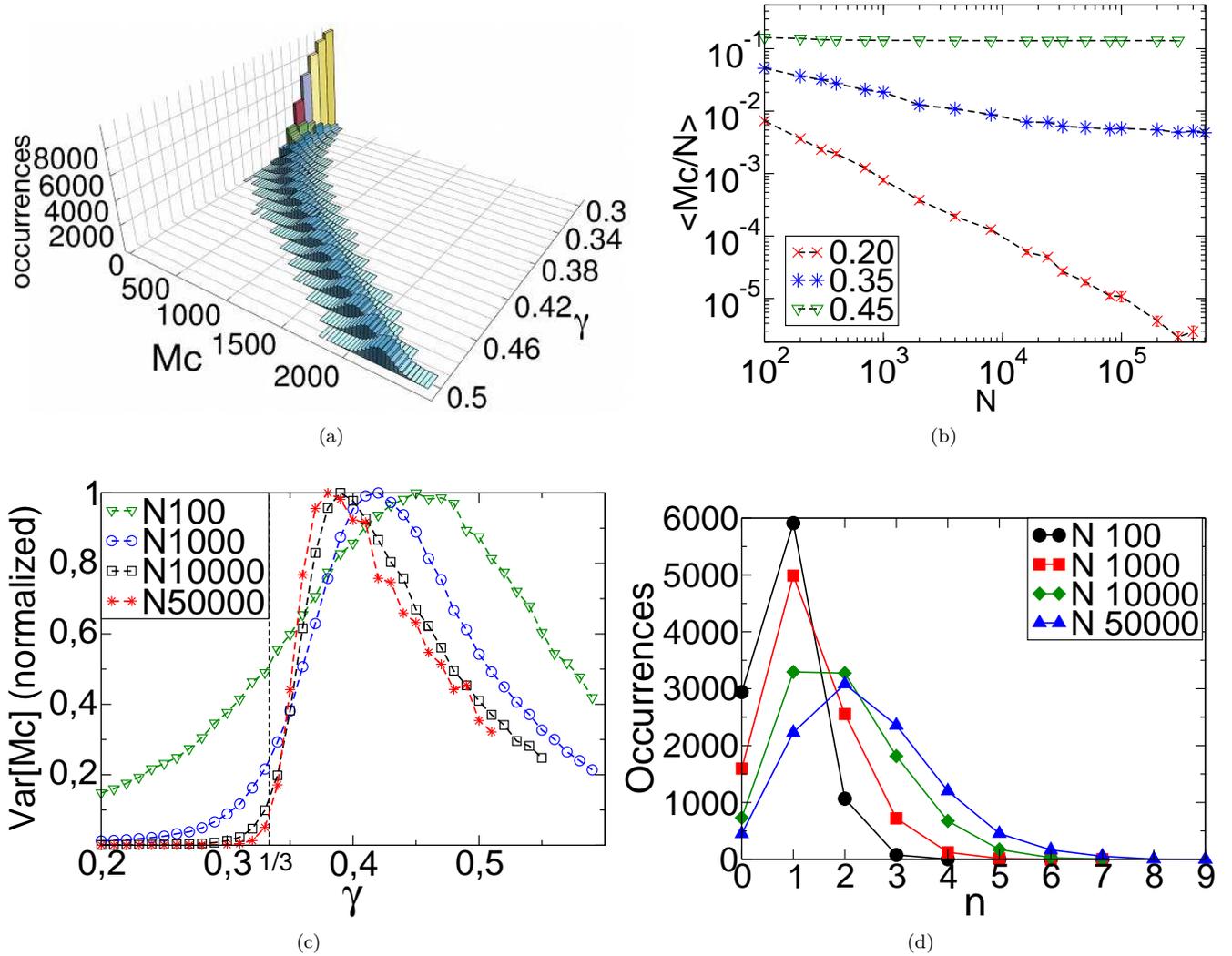}}\,
  \subfigure[\label{image_both:a}]%
  {\includegraphics[width=0.45\linewidth]{fig6_b.eps}}\\
  \subfigure[\label{FeedBack:b}]%
  {\includegraphics[width=0.495\linewidth]{fig6_c.eps}}
  \subfigure[\label{connected}]%
  {\includegraphics[width=0.49\linewidth]{fig6_d.eps}}
  \caption{\textbf{Feedback core.} (a) Three-dimensional plot showing the histogram of the LRb-core size 
  $Mc$ for different values of gamma  (for $k = 3$, $N = 10^4$, and $10^4$ different networks, colors online). 
  (b) Fraction of nodes in the LRb core. Points represent the averaged values resulting from
    simulations (ensembles composed by $10^4$ networks or $10^3$ for
    large systems) for different values of $\gamma$: 0.20 (red online), 0.35
    (blue online) and 0.45 (green online). (c) Variance of the fraction of the
    number of nodes belonging to the core. (d) Histogram of the number
    $n$ of connected components of the core at varying $N$. For fixed
    values of $N$ the number of connected parts does not remarkably
    change remaining around the critical value of $\gamma$
    (right). $10^4$ iterations for each value of $N$ are simulated.    
\label{FeedbackCore}}
\end{figure*}

We will now consider the topology of networks with $\gamma \simeq
\gamma_c$ (critical networks). Starting from Eqns.~
(\ref{prob_both}) we can write
\begin{eqnarray*}
	f_1  & = & \frac{1}{z^*_{up}} e^{-\lambda(z^*_{up})} \lambda(z^*_{up})  
\simeq  e^{-2k\epsilon}  
\simeq 1- 2k\epsilon + 
O(\epsilon^2) \\
	p_1 & \simeq & (1-\frac{2k\epsilon}{k-1})^{k-1}  \simeq 1- 2k\epsilon + 
O(\epsilon^2) \;,
\end{eqnarray*} 
concluding that $\frac{f_2}{f_1} \sim \epsilon$ and $\frac{p_2}{p_1}
\sim \epsilon$. This means that, near to the transition point and at
the thermodynamic limit, core matrices typically have one entry only
per-column \textit{and} per-row. In other words, they are permutations
of the identity matrix (and consequently they are invertible). Thus,
we can think of the residual (LRb) core as a particular Kauffman
network with in-degree one \cite{drossel:aa, flyvbjerg:1988aa}. There
is, however, an important difference. In Kauffman networks with
in-connectivity one, a node can have more than one outgoing edge,
while in this case all nodes in the core must regulate exactly one
element; there are no nodes without output (otherwise they would have
been removed by the algorithm).

\begin{figure*}[!htbp] 
  \centering%
  \subfigure[\label{core:a}]%
  {\includegraphics[width=0.5\linewidth]{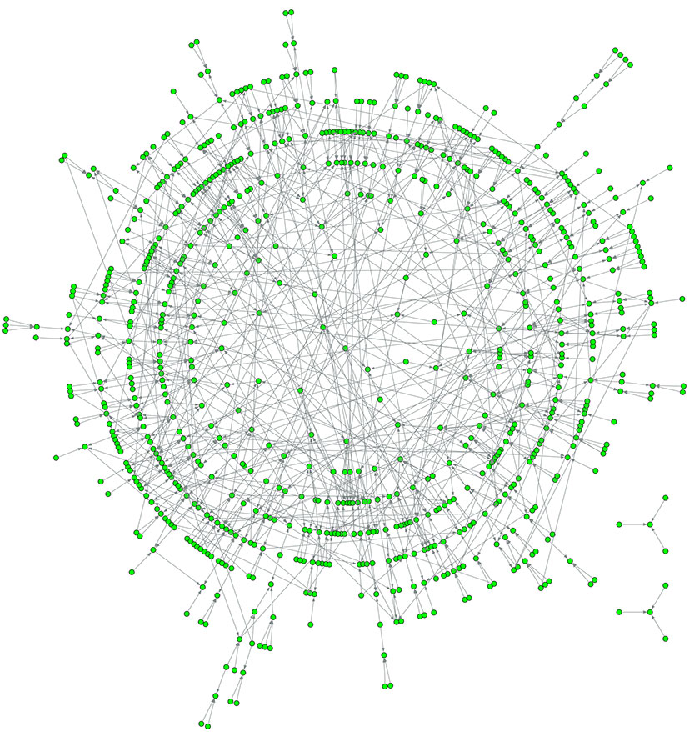}} \\
  \subfigure[\label{core:b}]%
  {\includegraphics[width=0.2\linewidth]{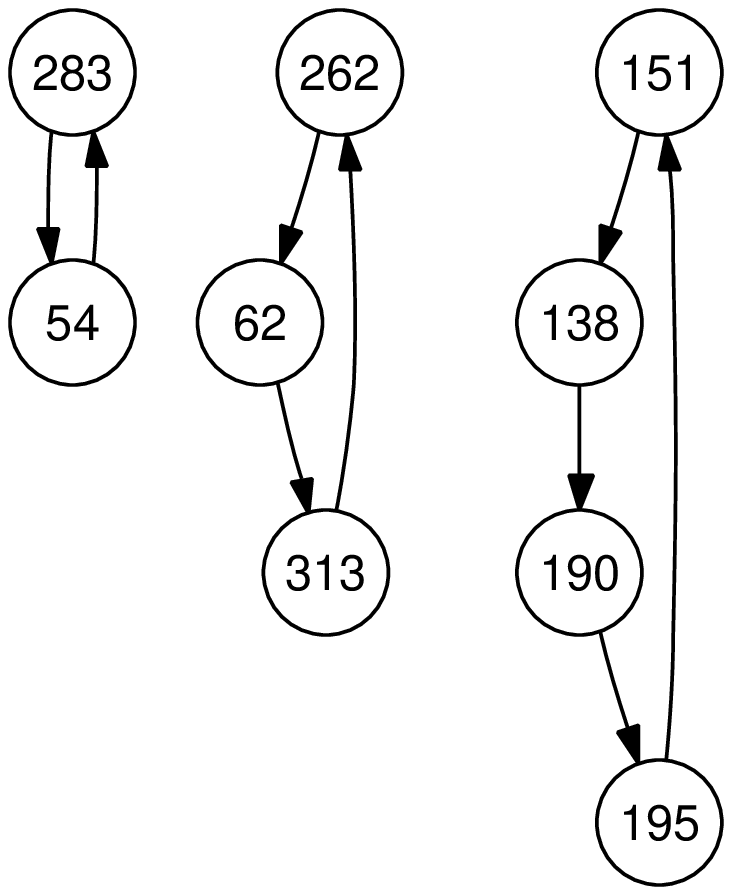}}\quad\quad
  \subfigure[\label{core:c}]%
  {\includegraphics[width=0.24\linewidth]{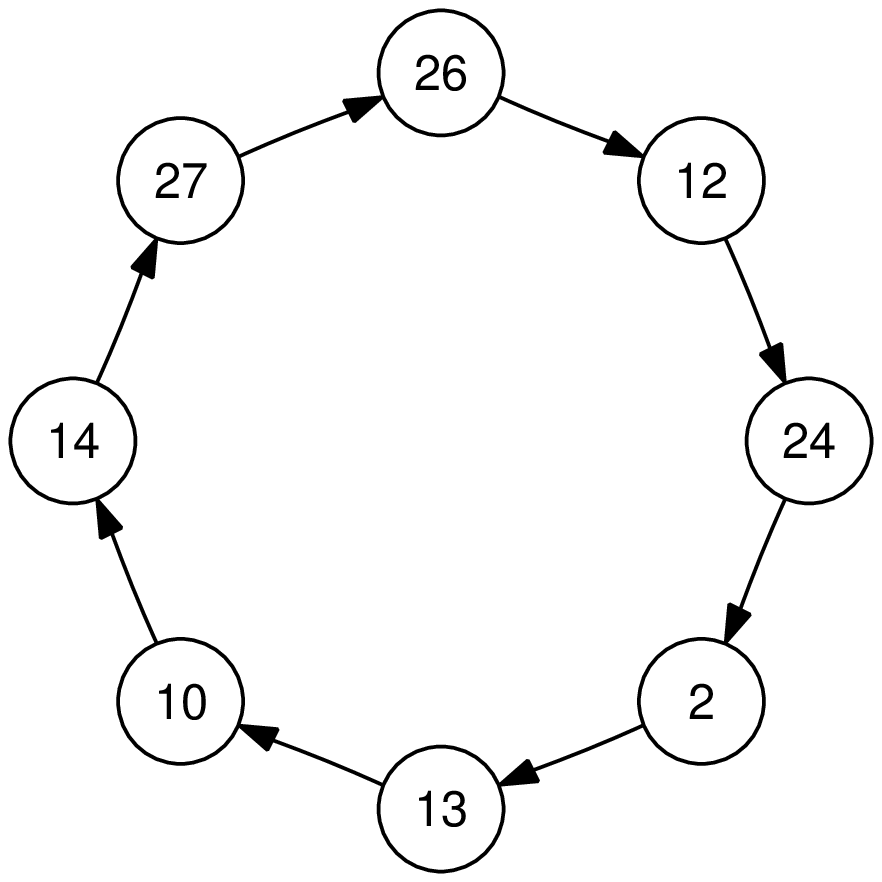}}
  \caption{\textbf{Structure of a critical core.} (a)
    shows a realization of a simulated network with
    $N=10^3$ and $M=333$ (potential isolated nodes are
    not depicted). It can be noticed that each regulated
    node receives exactly $k=3$ inputs. The structure is
    highly simplified after the application of LRb: only
    three disjoint simple loops are present (b). Panel
    (c) reports another typical chain structure of a
    core made of a single connected component (the
    starting graph had $N=100$ and $M=33$).\label{core}}
\end{figure*}

Having one entry per column and per row, core matrices correspond to
graphs in which all nodes have (typically) one input and one output,
and they show a structure with simple loops disconnected from each
other (Figs.~\ref{core}.b and \ref{core}.c).  Simulations suggest the
presence of several disconnected components that increase in number
when the dimension $N$ of the system increases. With growing $N$, the
number of tree-like graphs, as well as the number of cores with only
one connected component, decreases (Fig.~\ref{FeedbackCore}.d).  The
organization in disconnected loops does not depend on the
in-connectivity degree $k$. The number of nodes in the core of
critical networks scales as $N^{\zeta}$ with $\zeta \simeq0.4$
(Appendix~\ref{sec:appB}).

\subsection{Connecting feedback topology and dynamics: reduced 
dynamics 
\label{sec:dynamic_core}}

\begin{figure}[htbp!]
	\centering 
	\includegraphics[width=0.65\linewidth]{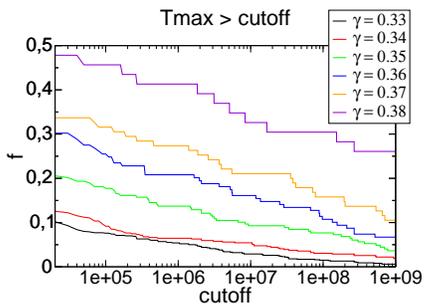}  
	\caption[Cycle time increases rapidly when $\gamma \gg
        \gamma_c$]{Fraction $f$ of networks which have a $T_{max}$
          greater than a cutoff. Simulations made for networks with
          $N=999$ and $k=3$. From $\gamma=0.33$ to $\gamma=0.38$ we
          respectively simulated 1000, 589, 284, 149, 95, 46 different
          realizations of the model.	\label{pass_cutoff}}
\end{figure}

Since topology and \textsc{xor} dynamics are strictly related, one
would expect that the feedback topology transitions have a
repercussion on the evolution of states. We first observe that, around the 
transition point, the matrices 
are invertible, hence $d=0$ 
(Section~\ref{sec:dyn}). From this, the probability of having at least one 
fixed points falls down at the 
transition. On the other hand, 
before the transition one expects to have only fixed points, because of the 
tree-like structure of graphs.
We can rearrange the indices
of the matrix $S$ and place \textit{leaf} variables (erased by LRu) in
the first indices and \textit{roots} (erased by LRd) in the lasts. In
this way it takes the form \cite{cosentino-lagomarsino:2006aa}:
\begin{equation*}
S =%
\left(\begin{array}{c}\mathcal{L} \\ \hline
\begin{array}{cc}0 \quad \phantom{a} & \begin{array}{|ccc} &  &  \\ 
\phantom{AB} & \mathcal{C} &  \\ &  & 
\end{array}  \end{array} \\ 
\hline 
\begin{array}{cc} \begin{array}{cccc} \phantom{a} & \phantom{a} & 0 & 
\phantom{a} \end{array} & 
\begin{array}{c|c}\phantom{a} & \: 
\mathcal{R}\end{array}\end{array}\end{array}\right) \;.
\end{equation*}
Nodes belonging to the LRb core are collected in the sub-matrix
$\mathcal{C}$ which has $Mz^*_{both}$ rows.  Applying $S$ to a vector
of the form $\Vec{x}(t) = (\Vec{l}(t) \, , \, \Vec{c}(t) \, , \,
\Vec{r}(t))$ (see Eq.~(\ref{x(t)}) ) one concludes that the evolved
state of roots, $\Vec{r}(t)$, is fixed only by initial conditions and
remains constant.  The nodes $\Vec{c}$ belonging to the LRb core are
determined by both $\Vec{r}$ and $\Vec{c}$ itself. Finally, the
configuration of leaf nodes $\Vec{l}$ is established by all the
variables and is not affected by feedback.  Thus, the behaviour of the
whole dynamics can be deduced from the state of the core. For these
reasons, simulations of the dynamics restricted to the core
(\textit{reduced dynamics}) are sufficient to evince the salient
features of the dynamics. Using this fact, we are able to simulate
networks with up to $10^3$ nodes with a large statistic. On the other hand, 
obtaining data
in the chaotic region is very difficult also using the reduced
dynamics (see Fig.~\ref{pass_cutoff}).

Nevertheless, the region immediately after $\gamma_c$ is interesting,
because it carries the consequences of the simplified core structure
in the topology.  The equations for critical core variables are
special cases of Eq.~(\ref{xor}) when each node $i$ receives one
input from node $j(i)$:
\begin{equation*}
\label{xor_trans}
c_i (t+1) = c_{j(i)}(t) + b_i \;,
\end{equation*}

In case the core is a single component, then the maximum period length
$T_{max}$ will be approximately equal to $M_c$.  Since, near the
transition, core matrices are invertible Eq.~(\ref{cycl}) becomes
$S_{c}^{t_{m}} = 1$ ($l=0$).  The condition to have the maximum period
is
\begin{equation*}
\Vec{c}(T_{max}) = S_{c}^{T_{max}} \Vec{c}(0) + \Sigma(T_{max})\Vec{b} = 
\Vec{c}(0) \;.
\end{equation*}
The matrices $S_c$ are also permutation matrices with dimension $M_c
\times M_c$ and thus $S_{c}^{M_c} = 1$. Since $\Sigma(2M_c)=0$, the
maximum period achievable is $T_{max}= 2M_c \sim M_c$.  In case the
core is formed by several disconnected chains, the maximum period of
the whole network can be computed as the least common multiple of the
period of each single loop (Appendix~\ref{sec:appC}).

\section{Discussion and conclusions}
In this work, we considered a model of Random Boolean Networks related
to Kauffman networks. The model has the objective to investigate on
abstract grounds some features that are important in biological
networks, mainly (1) the presence of nodes which regulate but are not
regulated, existing in empirical transcription networks, and (2) the
correspondence between topology and dynamics. For the latter reason we
chose to use a \textsc{xor} dynamics to define the interaction between
elements. We have shown that in this model the dynamical aspects
characterizing the length of cycles and their basins of attraction are
direct consequences of the topological feedback structure of the
underlying networks, which we study with graph decimation algorithms
(LRs) on a fixed in-degree ensemble of graphs.\\
\indent We identified a dynamical and topological phase transition with
varying input structure of the graphs modulated by the parameter
$\gamma$ (the fraction of the input-receiving nodes).  Direct
numerical simulations of the dynamics (Figs.~\ref{old} and
\ref{over_cf}) suggest the presence of a phase transition where long
cycles emerge.  Correspondingly, a topological phase transition
separates a tree-like graph region from one in which extensive
feedback components are present.  Different Leaf Removal algorithms
that remove the upstream tree-like regions, the downstream ones, or
both, have the same critical point $\gamma_c = k^{-1}$, which we
locate analytically and numerically.
The dynamic transition is found numerically at a critical value that
is close to $\gamma_c$, but strongly influenced by finite size
effects.
Since networks below the transition typically exhibit fixed points,
and this behaviour is correlated with the tree-like structure of these
graphs, we identify the dynamic transition point with $\gamma_c$ in
the thermodynamic \nopagebreak limit.
Thus, despite of the large finite-size effects which make the
simulations difficult, our results lead to the conclusion that the
topological and dynamic transitions are the same.\\
\indent This result is only valid for the ensemble of functions under
consideration. Dynamics with a more general class of functions should
present a different critical point. However, the value of the nodes
removed by LRd is fixed after a transient time, independently from the
dynamics chosen, because the fact that nodes receiving inputs from the
topological core depend just on the the state of the core is not
conditioned to the class of functions.  The relevant dynamical part of
the network must be a subset of this the core, and a different choice
of update rules moves the dynamical transition away from the
$\gamma_c$.
It is simple to imagine that with a more general class of functions
one can still observe fixed points for values of $\gamma \gtrsim
\gamma_c$. Thus, the topological critical point represents a lower
bound for the dynamical critical point.\\
\indent Above the transition (Fig.~\ref{over_cf}) the presence of an
extensive feedback region induces chaotic dynamics, characterized by
exponential cycles (order $2^{M_c}$ with $M_c \sim O(N)$). In this
case, the initial conditions, and thus the influence of the external
world (represented by the state of sensor nodes in the fraction
$1-\gamma $), do not affect the dynamics.  For this reason, studying
the features of
networks away from the transition is only relatively interesting.\\
\indent The most interesting region to study is the transition point, where
the dynamics can be at the same time nontrivial and under external
control. At this point, our analytical mean field theory predicts that
the structure of feedback components in the graph is simple.  Feedback
components (independently from the in-degree) show an elementary
modular organization in simple disconnected loops, found in
simulations (Fig.~\ref{core}) and predicted by the analytical
approach.  In turn, this feedback structure transposes to a modular
dynamics, which can be completely characterized by the \textsc{lcm}
structure of periods. We give an analytical (large $N$) prediction for
the maximum cycle achievable. Comparing with (small $N$) simulations
(Fig.~\ref{Tmax}), the distribution of the maximum cycles is wide but,
with increasing dimension of the system, the estimate becomes
increasingly reliable and comparable with the prediction. This point
might have an interest for the modeling of empirical biological
networks (which have $N$ of the order of a few thousand).  Indeed, at
these ``realistic'' system sizes, this point corresponds to an
interval with a finite span, because of finite-size effects, as it
happens with Kauffman networks \cite{socolar:2003aa}.\\
\indent It is interesting to make a comparison between the model used here and
the behaviour of Kauffman networks with no input structure in the
underlying graph and general Boolean functions. In the latter model,
the dynamics is entirely controlled by so-called relevant nodes
\cite{flyvbjerg:1988aa, bilke:2001aa} which have been the subject of
many studies \cite{socolar:2003aa, kaufman:2006aa, krawitz:036115,
  bastolla:1998aa}.  In networks between the chaotic and the stable
phase, they spontaneously organize into disconnected clusters whose
number increases logarithmically with the system size
\cite{bastolla:1998ab, kaufman:2006aa}. Most of relevant clusters are
simple loops. The number of relevant nodes in critical networks scales
as $N^{1/3}$ in the limit $N \rightarrow \infty$
\cite{socolar:2003aa}. As recently found by Drossel and coworkers
\cite{kaufman:2006aa, kaufman:2005ab, mihaljev:2006aa}, since the
relevant part constitutes only a vanishing portion of the network, the
topology of Kauffman networks is most likely very different from the
topology of real biological networks, such as genetic regulatory
networks, where one would expect that the majority of nodes is
relevant or at least not always frozen.\\
\indent As anticipated in Section~\ref{sec:model}, there exists an analogy
between relavant components of Kauffman networks and the LRb core of
the $\gamma$ model. We find that the critical core has a modular
structure and the amount of disconnected loops increases as the system
size grows.\\
\indent Considering the scaling approach of Mihaljev and coworkers
\cite{mihaljev:2006aa}, it is interesting to observe the close
correspondence between an \textsc{xor} Kauffman network with an
imposed fraction of $1-\beta$ of constant functions and the model
studied here.  In this case, their approach would give a transition
between frozen and chaotic phase located at $\beta_c = 1/K$ in the
parameter space. Note that, however (since no \textsc{xor} functions
are constant) the only way to vary $\beta$ in our case is to change
the \emph{ensemble} of graphs, i.e. vary $\gamma$, and allow for a
input-output structure of the network.
The difference between the two approaches is acting directly on the
topology (more specifically, on the fraction of input nodes) instead
of modifying the graph with the introduction of constant functions.
The same reasoning suggest that this transition point we find is a
lower bound for Kauffman networks with a fraction $1-\gamma$ of
constant functions and a general ensemble of functions.\\
\indent In spite of this analogy, simulations show that the number
$M_c$ of nodes belonging to the core of critical networks scales as
$N^{\zeta}$ with $\zeta \simeq 0.4$ (see Appendix~\ref{sec:appB}) that
is different from the scaling exponent found for relevant components
in critical Kauffman networks (discussed for example in the same study
of Mihaljev and coworkers, \cite{mihaljev:2006aa}).
The error on the fit seems to be small enough to exclude a discrepancy 
of $0.1$ in the exponent but it is possible that this deviation between the two models 
derives from the fact that we have accessed this quantity by numerical
work, and thus are limited by system size (we tested up to 250.000 nodes).
On the other hand, we can also speculate that this difference derives
from the fact that the topology of the underlying graphs in the two
models is not the same. Indeed, in the case of a general Kauffman
network, nodes may become frozen because some of their inputs are
connected to a frozen node, and the resulting function is constant.
Thus, frozen nodes do not have strictly zero in-degree as in the
$\gamma$ model. In our case, this cannot happen, as the topology is
strictly controlled.  In other words, the quantity $M_c$ describes the
size of the feedback region that \emph{only for the $\gamma$ model 
with \textsc{xor} dynamics} coincides with the relevant dynamic
region.
Studying the simplest possible case of this model has the further
advantage of elucidating in detail some of the phenomenology connected
to the properties of the critical core, its modular structure and the
amount of disconnected loops, and thus providing an estimate for the
size of the largest cycle in the dynamics.\\
\indent In conclusion, while simplified, the $\gamma$ model gives an insight
into the role of the input-output structure in the whole-network
dynamics, which is relevant for gene networks. Direct application to
concrete problems concerning regulatory networks would require
generalizing the approach to more realistic representations for the
input functions and graph ensembles \cite{bassetti:2007aa}. In
particular, for what concerns the dynamics, our results on the role of
the emerging extensive feedback core, which apply to \textsc{xor}
functions, could be extended to the study of more complex dynamics
with more realistic choice for the ensemble of functions.
As soon as a different classes of Boolean functions are included in
the model, the situation becomes more complicated, as the feedback
structures responsible for the dynamical behaviour of the network would
not anymore simply be the cycles in the network's topology, but a
subset of them. On the other hand, it is possible that this problem
can be bypassed by defining generalized pruning algorithms on the
underlying graphs where different ``weights'' are associated to each
link or Boolean gate.  Thus, the basic phenomenology and tools
described here could be exploited in constructing and approaching
simplified but realistic models of genetic regulation networks.
\appendix
\section{Leaf Removal equations \label{sec:appA}}
\subsection{Leaf Removal up}
To write the mean-field equations for this algorithm, we analyze the
evolution of connectivities of the regulated nodes only, which are
specified by the matrix $S$.  The number of LR steps is the time $t$
of the process and is the number of nodes that have been removed.  We
introduce the \textit{normalized time} $\bar{t} = t/M$ , $\bar{t} \in
[0, 1]$ so that $\Delta \bar{t} = M^{-1}$ and $\Delta c_n = M \Delta
f_n = \frac{ \partial f_n} { \partial \bar t}$.

\noindent At time $t=0$, the probabilities $f_n$ (see Section~
\ref{sec:LeafRemoval}) are 
\begin{equation}
  \begin{cases}
   	f_{X}(0) = 0 \\
	f_{n}(0)  = e^{-\gamma k}\frac{(\gamma k)^n}{n!} \doteq e^{- 
\lambda(0)}\frac{\lambda (0)^n}{n!} \; , 
\qquad  n\geq 0 \;,
   \label{fn0}
   \end{cases}
\end{equation}
$f_{X}$ indicating the fraction of erased nodes.
\noindent When $M f_{X}(\bar t \,) = M \bar t$ nodes are removed, 
$c_n(\bar t \,)  = M f_n(\bar t \,)$ 
nodes with $n$ outgoing edges 
remain.
When a node without outgoing edges is found, it is removed (it 
corresponds to a column of the matrix 
with no elements different 
from 0) so that at each step $Mf_{X}$ increases of one unit.
Also its incoming links are removed, which means that the corresponding 
row in $S$ is cleared out, 
replacing with zero all the 
entries. 
Removing edges changes the probability distribution $f_n(\bar t \,)$.
\noindent Indeed the probability that an edge that is being removed comes 
from a node with $n$ 
outgoing edges (this one 
becoming an $n-1$ outgoing edges node) is
\begin{equation*}
	P_{[n \rightarrow n-1]} = \frac{n c_n(\bar t \,)}{\sum_{n\geq 1} n 
c_n(\bar t \,)} = \frac{n f_n(\bar t \,)}
{\sum_{n\geq 1} n f_n(\bar t 
\,)} \;.
\end{equation*}
so that removing an edge implies $ \Delta c_n = -1\cdot P_{[n \rightarrow 
n-1]} + 1 \cdot  P_{[n+1 
\rightarrow n]} $.

\noindent Recognizing that we remove, on average, $\gamma k$ edges 
per step \footnote{This is the 
$in$-connectivity which 
remains constant as removal does not affect $in$-distribution $p_n$. The 
reverse will be true for LRd},
we write the flow equations for the probabilities $f_n(\bar t \,)$ as follows:
\begin{equation}
   \label{LRup_flux}
   \begin{cases}
   	\frac{ \partial} { \partial \bar t}  f_{X}(\bar t\,) = 1 \\
	\frac{\partial}{\partial \bar t} \ f_0(\bar t\,) = -1 + \frac{\gamma k}{\langle 
n \rangle (\bar t\,)}f_1(\bar t
\,)\\
    \qquad \qquad \: \vdots \\
    \frac{\partial}{\partial \bar t} \ f_n(\bar t\,) = \frac{\gamma k}{\langle n 
\rangle (\bar t\,)}[(n+1)f_{n+1}(\bar 
t\,)- nf_n(\bar t\,)] \:, n>0\\
   \end{cases}
\end{equation}
with initial conditions (\ref{fn0}) and where $ \langle n \rangle (\bar t\,)  = 
\sum_{n\geq 1} n f_n(t) = 
\sum_{n\geq 0} n f_n(t) = k 
\gamma (1- \bar t\,)$ being the average number of edges per node after 
$M \bar t$ removals.
One can easily check that the poissonian $f_{n}(t)$ given in Section~
\ref{sec:LeafRemoval} are 
solutions of 
Eqns.~(\ref{LRup_flux}) once having imposed
\begin{equation*}
	-\frac{ \frac{d\lambda(\bar t\,)}{dt} }{\lambda(\bar t\,)} = \frac{\gamma k}
{\langle n \rangle (\bar t\,)} = 
\frac{1}{1-\bar t} \;, 
\end{equation*}
from which $\lambda(\bar t\,) = \gamma k (1- \bar t\,)$. 
See Ref.~\cite{cosentino-lagomarsino:2006aa} for details.
The algorithm stops when $f_0(\bar t\,) = 0$, i.e. at time $\bar t^* =  e^{- 
\lambda(\bar t^*)}$. This 
equation implies that if $\gamma k 
< 1$ the stop normalized time is $\bar t^* = 1$ or, in other words, all the 
graph is removed. There is a 
critical value of $\gamma$, $
\gamma_c = k^{-1}$, below which the whole graph is cancelled after the 
application of LRu. These 
networks typically have a tree-
like structure without feedback regions.\\ 
Being $t$ the number of variables deleted, it is useful to introduce the 
variable
\begin{equation*}
	z_{up} \doteq 1- \bar t = 1 - \frac{t}{M} = \frac{M-t}{M} \;.
\end{equation*}
which expresses the fraction of remained nodes during the LRu process .

\subsection{Leaf Removal down}
In the first LRd step, all the $N-M$ unregulated nodes are eliminated from 
the network because they 
have no inputs (this 
corresponds to neglecting the matrix $R$ and working with only the 
square matrix $S$). 

As above, the time $t$ of the algorithm represents the number of removed 
nodes.
It is again useful to employ a normalized time $\bar t$. 
The number of nodes with $n$ entries at time $t$ is $r_n(t) = M p_n(t)$ 
where $p_n(t)$ is 
the probability introduced in Section~\ref{sec:LeafRemoval}. 
At time $t=0$ it reads:
\begin{equation*}
	 \begin{cases}
	p_{X}(0) = 0 \\
	p_{n}(0) =\left(\begin{array}{c}k \\ n\end{array}\right) \gamma^n (1-
\gamma)^{k-n} \quad , \quad 
0\leq n \leq k \; ,
	\end{cases}
\end{equation*}
where $p_{X}$ is the fraction of removed nodes.
In the decimation algorithm a node with no entry is cleared out together 
with its outgoing connections, 
so that probability $p_n(t)$ changes. At each step, the number of emptied 
rows $r_{X}$ increases by 
one and we obtain
\begin{equation*}
    p_{X}(\bar t\,) = \bar t \qquad \qquad \sum_{i \geq 0}p_i(\bar t\,) = 1- \bar t 
\;.
\end{equation*}
The probability that a removed edge was input for a node with $n$ ingoing 
edges is:
\begin{equation*}
	P_{(n\rightarrow n-1)} =  \frac{n r_n(t)}{\sum_{n\geq 1} n r_n(t)} = 
\frac{n p_n(\bar t\,)}{k\gamma(1- 
\bar t \,)} \; 
\end{equation*} 
where we remembered that $ \langle n \rangle (\bar t\,) = k \gamma (1- \bar 
t\,)$.

\noindent Once again, an average number $k \gamma $ of edges are 
removed at each step so that
a node with $n$ entries becomes, with probability $k\gamma \cdot P_{(n
\rightarrow n-1)}$, a node with 
$n-1$ entries. 
Thus, one can write the flow equations:
\begin{equation*}
   \begin{cases}
   	\frac{\partial}{\partial z} \ p_{X}(z) = -1\\
	\frac{\partial}{\partial z} \ p_0(\bar t\,) = 1 - \frac{p_1(z)}{z}\\
	\qquad \qquad \: \vdots \\
	\frac{\partial}{\partial z} \ p_n(z)  =  n \frac{p_n(z)}{z} - (n+1) \frac{p_{n
+1}(z)}{z} \qquad n\leqslant 
k,\\
	 p_{k+1} = 0 \;, 
   \end{cases}
\end{equation*}
where we have already made the substitution $z = z_{down} = 1- \bar t$ ,  
$z_{down} \in [0,1]$. In 
Section \ref{sec:LeafRemoval} 
we give the solutions to these equations.
The process ends when no more unregulated nodes are available, that is 
when $p_0(z) = 0$, namely 
when $1-z = (1-\gamma z)^k 
$. Studying this condition one sees that if  $\gamma < k^{-1}$ then there 
are no solutions apart from the 
value $z=0$, which 
indicates the existence of a tree-like graph in which all is removed.

\section{Finite size effects	\label{sec:appB}}
According to our simulations, finite size effects appear to be very relevant 
both in topological structure 
and dynamic behaviour. 
Increasing $N$, the shape of the distribution remains the same and it is 
more peaked around the 
central value predicted by LR 
equations.
\begin{figure}[htbp]
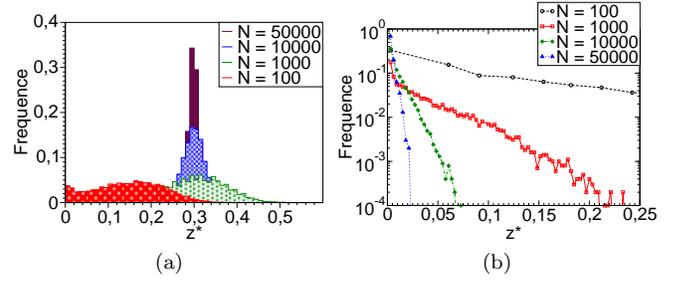
 
	\centering%
	\subfigure[\label{Mc_finite:a}]%
		{\includegraphics[width=0.485\linewidth]{fig9_a.eps}}
	\subfigure[\label{Mc_finite:b}]%
		{\includegraphics[width=0.499\linewidth]{fig9_b.eps}}
    	\caption{Distributions of the number of nodes belonging to the LRb 
core for different system 
dimension at fixed $\gamma$. 
General case (a) with $\gamma = 0.45$ and critical networks (b). 
Simulations are produced by $10^3$ 
iterations for $N=5 \cdot 
10^4$ and $10^4$ iterations otherwise. \label{Mc_finite}} 
\end{figure}
Furthermore, it can be observed that $M_c$ scales with $N$ as a power 
law with exponent lower than 
1 (simulations suggest a 
trend $M_c \sim N^\zeta$ with $\zeta \simeq 0.4$), while nodes upstream 
the core that are removed by 
LRd and nodes 
downstream removed by LRu grow approximately linearly in $N$. Figures 
\ref{scaling} show the 
results of simulations. The 
upstream part of the graph removed by LRd is indicated with the letter $U$ 
and the downstream part 
with $D$. One speculates that 
$U$, $D$ and $M_c$ at the transition point behave as
\begin{equation*}
	U \sim \alpha N \qquad
	D \sim \beta N \qquad
	M_c \sim N^\zeta \;.
\end{equation*}
Also the particular core structure of critical networks is affected by finite 
size effects. In fact, for $N=100$ 
it is not unusual to find 
core networks with more complex structures than disconnected simple 
loops. This also explains the 
behaviour of Fig.~
\ref{connected}: one can speculate from the plots that the LRb core 
description in modular structures 
becomes more accurate 
increasing $N$, while, for smaller systems and $\gamma_c \leqslant 
\gamma \leqslant \gamma_e$, a 
single disconnected 
component with a more complex structure is present with more frequency.  
 \begin{figure}[htbp!] 
	\centering
	 \includegraphics[width=0.75\linewidth]{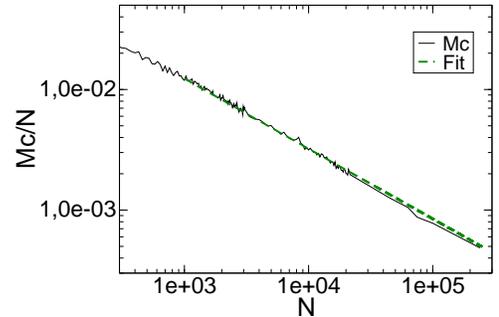} 
	\caption{Scaling with $N$ of the core $M_c$. Each point of simulation 
is the averaged result of 
$10^3$ iterations. The fit $y=a 
x^b$ with $a\simeq0.68$ and $b\simeq -0.58$ is presented.\label{scaling}} 
\end{figure}

\section{Estimates of $T_{max}$\label{sec:appC}}
\begin{figure*}[htbp]
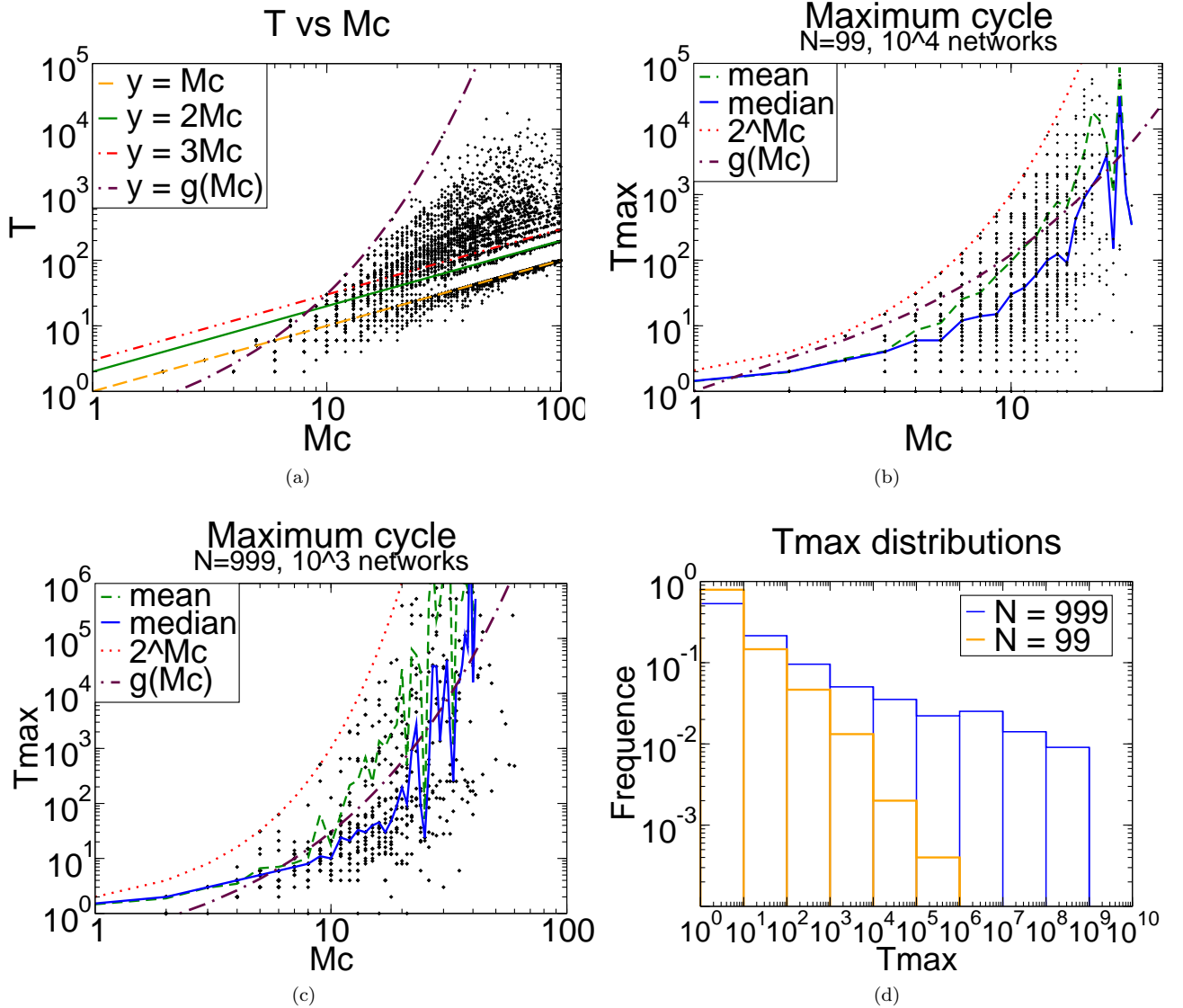
 
  \centering%
  \subfigure[\label{Tmax:a}]%
  {\includegraphics[width=0.49\linewidth]{fig11_a.eps}}
  \subfigure[\label{Tmax:b}]%
  {\includegraphics[width=0.47\linewidth]{fig11_b.eps}} \\
  \subfigure[\label{Tmax:c}]%
  {\includegraphics[width=0.49\linewidth]{fig11_c.eps}}
  \subfigure[\label{Tmax:d}]%
  {\includegraphics[width=0.46\linewidth]{fig11_d.eps}}
  \caption {\textbf{Maximum cycle length for critical networks.} (a)
    Values of $T = \mathrm{lcm}\{l_i\}$ vs $Mc$ for $\gamma=0.3334$,
    $N=10^4$, and $10^4$ different networks. $g(M_c) \sim \exp
    (\sqrt{M_c \ln M_c})$. (b) and (c) are numerical simulation of the
    dynamics of core networks at $\gamma =0.33$ ($k=3$). The plots
    report the values of $T_{max}$ as a function of $M_c$. Dots are
    experimental points compared with the bound $g(M_c) \sim \exp
    (\sqrt{M_c \ln M_c})$. Solid line (blue online) reports the median and the 
dashed one (green online) 
the mean. The
    cutoff on cycle lengths is $10^9$. For $N=999$ (c) this limit is
    passed only six times and all for $M_c \geqslant 48$. Estimates of
    central values are not calculated when data are insufficient. (d)
    Shape of the distributions of $T_{max}$ for $N=99$ (red online) and
    $N=999$ (blue online). The distributions are power-law-like and their
    tails are limited by the cutoff imposed ($10^9$). \label{Tmax} }
\end{figure*}

We estimate $T_{max}$ considering cores built of
\textit{loop} structures \footnote{we call \textit{loop} the topological
  chains reserving the term \textit{cycle} to the dynamics} of lengths
the first $m$ prime numbers $\{p_1, p_2, \ldots ,p_m\}$, so:
\begin{eqnarray*}
	M_c = \sum_{i\leqslant m} p_i \: , \qquad T = T_{max} =\prod_{i \leqslant m} p_i \;.
\end{eqnarray*}
Since the $n$th prime number scales as $n \ln n$, from the above 
equations one has
\begin{eqnarray*}
  M_c & \sim & \sum_{n}^{m} n\ln n \sim \frac{1}{2}m^2 \ln m \\
 \ln T_{max} & \sim & \sum_{n}^{m} \ln n +\ln \ln n \sim m \ln m  \;,
\end{eqnarray*}
from which 
\begin{equation}
  \ln T_{max} \sim \sqrt{M_c \ln M_c} \;.
  \label{lnTmax}
\end{equation}
Simulated graphs show that this expression is a good upper bound for
the value of $T = \textsc{lcm}\{l_i\}$ (Fig.~\ref{Tmax}.a). The
plots indicate that these values ``thicken'' around multiple values of
$M_c$. This suggests that residual graphs have a large component with
dimension roughly equal to $M_c$ and a short loop of length 2, 3,$\ldots$\\
It is interesting to compare these estimates with maximum-length
cycles generated by simulations of the core dynamics. Our numerical
plots show the values of the maximum period $T_{max}$ emerging from
reduced dynamics. In the case $N=99$ the bound set by 
Eq.~(\ref{lnTmax}) 
is passed several times (Fig.~\ref{Tmax}.b) because of
finite size effects, but it becomes more reliable for larger
systems. Figure \ref{Tmax}.c shows that simulated values for $T_{max}$ 
seem
to concentrate under the function $g(M_{c}) \sim \exp (\sqrt{M_c \ln
  M_c})$.
Finally, Fig.~\ref{Tmax}.d shows the maximum cycle distributions for the
same simulations. These are power-law like, and become broader with
the system dimension. Since statistics is limited by the imposed
cutoff, averages are highly biased by this computational restriction
and cannot be considered reliable for large systems or much beyond the
critical line.

\end{document}